\documentclass[pra, superscriptaddress, twocolumn, aps]{revtex4-1}

\usepackage{amsmath, amssymb, amsfonts}
\usepackage{dsfont}
\usepackage{bm}
\usepackage{graphicx}
\usepackage{subfigure}
\usepackage{hyperref}
\usepackage{mathrsfs}
\usepackage{yfonts}

\newcommand{\cohosub}[1]{\scalebox{0.72}{\textswab{#1}}}

\newcommand{\coho}[1]{\textswab{#1}}
\newcommand{\mb}[1]{\mathbf{#1}}
\newcommand{\Ref}[1]{Ref.~[\onlinecite{#1}]}
\newcommand{\Refs}[1]{Refs.~[\onlinecite{#1}]}

\begin{document}

\title{Classification of symmetry fractionalization in gapped $\mathbb Z_2$ spin liquids}
\author{Yang Qi}
\affiliation{Perimeter Institute for Theoretical Physics, Waterloo, Ontario
  N2L 2Y5, Canada}
\affiliation{Institute for Advanced Study, Tsinghua University,
  Beijing 100084, China}
\altaffiliation[Current address: ]{Department of Physics, Fudan University, Shanghai 200433, China}
\author{Meng Cheng}
\affiliation{Station Q, Microsoft Research, Santa Barbara, California 93106-6105, USA}
\altaffiliation[Current address: ]{Department of Physics, Yale University, New Haven, CT 06520-8120, USA}

\begin{abstract}
  In quantum spin liquids, fractional spinon excitations carry half-integer spins and other fractional quantum numbers of lattice and time-reversal symmetries. Different patterns of symmetry fractionalization distinguish different spin liquid phases. In this work, we derive a general constraint on the symmetry fractionalization of spinons in a gapped spin liquid, realized in a system with an odd number of spin-$\frac12$ per unit cell. In particular, when applied to kagome/triangular lattices, we obtain a complete classification of symmetric gapped $\mathbb Z_2$ spin liquids.
\end{abstract}

\maketitle

\section{Introduction}
\label{sec:introduction}

Gapped spin liquids~\cite{BalentsSLReview,SavaryBalentsReview} are strongly-correlated quantum states of spin systems that do not exhibit any Laudau-type symmetry-breaking orders~\cite{landau}, but instead are characterized by the so-called topological order~\cite{KivelsonPRB1987, WenTO1990, Wen1991a, WenZ2SL1991}. The existence of the topological order manifests as the emergence of quasiparticle excitations with anyonic braiding and exchange statistics. Although no symmetries are broken, the nontrivial interplay between the symmetries and the topological degrees of freedom further enriches the concept of topological orders, leading to the notion of symmetry-enriched topological (SET) order~\cite{XChenLUT}. In SET phases, global symmetries can have nontrivial actions on the quasiparticle excitations, which are otherwise not allowed on non-fractionalized local excitations (i.e. magnons). In particular, quasiparticle excitations can carry fractionalized symmetry quantum numbers, a phenomena known as symmetry fractionalization~\cite{Essin2013}.

An ubiquitous pattern of symmetry fractionalization in quantum spin liquids is that there exists a quasiparticle excitation, dubbed the ``spinon''~\cite{KivelsonPRB1987,WenZ2SL1991}, which carries a half-integer spin quantum number
~\footnote{Although the physical degrees of freedom in a frustrated magnet are often half-integer spins, all local excitations, like magnons, must carry integer spins, which are linear representations of the SO(3) spin-rotational symmetry group. Therefore, we should describe the physical spin-rotational symmetry with an SO(3) symmetry group, and treat a half-integer spin as a fractional symmetry quantum number, which labels a projective representation of the SO(3) symmetry group.}.
Besides the spin-rotational symmetry, quantum spin liquids often have time-reversal symmetry as well as crystalline symmetries of the underlying lattice. Correspondingly, the spinon can also have fractionalized quantum numbers of the space-time symmetries. This was first discussed in the framework of parton constructions of spin liquid state, known as the projective symmetry group (PSG) of the parton mean-field ansatz~\cite{wenpsg}. In this work, we will use spinon PSG to collectively refer to space-time symmetry fractionalization on spinons. Together with other fractionalized symmetry quantum numbers, PSG classifies symmetric quantum spin liquids that can be realized on a certain type of lattice.

Classification of symmetry fractionalization is not only of central importance to the understanding of SET phases, but also essential to the studies of quantum spin liquids. Spin liquids with different patterns of symmetry fractionalization belong to different phases, and thus are represented by different symmetric variational wave functions~\cite{sstri,wenpsg,Wang2007,YMLuKagomePSG2011,SJiangTPS2015X}. Hence, a complete classification exhausting all possible universality classes of topological spin liquids greatly facilitates the numerical studies of such exotic states. Furthermore, symmetry fractionalization may lead to interesting physical signatures~\cite{wenpsg, EssinPRB2014, WangPRB2015, ZLVPSG, QiCSF}, which can then be detected both numerically and experimentally. Therefore, symmetry fractionalization provides an important set of features one can use to study the seemingly featureless quantum spin liquid states.

Viewing symmetric gapped quantum spin liquids as SET states, it has been realized that PSGs of different parton constructions can be unified~\cite{LuBFU}, and together they classify a subset of possible SET states~\cite{EssinPRB2014}. In addition to the spinon PSG, the SET states are also characterized by fractionalized quantum numbers of the vison excitations, which we will briefly refer to as the vison PSG, and possibly an additional layer of a symmetry-protected topological (SPT) state~\cite{XChenSPT} (in this work, we will ignore this further distinction, i.e., we regard two SET states related by stacking a SPT layer as the same state). Naively, taking vison PSGs into account would significantly enlarge the classification table~\cite{EssinPRB2014}. However, it turns out that many SETs where both spinons and visons exhibit nontrivial symmetry fractionalization are anomalous~\cite{VishwanathPRX2013, CWangETMT2013, Qi_unpub, HermeleFFAT, QiSFV2015X,Song_unpub}: they can not arise in truly two-dimensional lattice systems, and can only be realized on the surface of a three-dimensional SPT state. For on-site unitary symmetries, whether a particular symmetry fractionalization pattern is anomalous or not can be determined mathematically~\cite{Chen2014,BarkeshliX}. Unfortunately, such a mathematical framework does not yet exist for space-time symmetries. Instead, in this work, with the assumption of a background spinon charge per unit cell, we are able to obtain a set of nontrivial constraints on non-anomalous 2D spin liquids.

More specifically, we derive constraints on spinon's symmetry fractionalization in gapped $\mathbb Z_2$ spin liquids with a background spinon charge per unit cell, using the triangular/kagome lattice (these two lattices have the same crystalline symmetry group) as the main example. We show that in these systems
the spinon PSG must match the projective
representation of the physical degrees of freedom in a unit
cell. Similar constraints are introduced in \citet{Cheng_unpub} for onsite symmetries. In this work, we show how to generalize this type of constraints to include point-group symmetries. Together with the constraints on vison's symmetry fractionalization derived in \Ref{QiSFV2015X}, we have for the first time obtained a full classification (up to stacking SPT layers) of symmetric gapped $\mathbb Z_2$ spin liquids on a triangular/kagome lattice in two dimensions.

We will focus on possible gapped $\mathbb{Z}_2$ spin liquids in spin-$\frac12$ antiferromagnetic(AF) Heisenberg models on the triangular and kagome lattices, because they have become the most promising candidate systems to realize these exotic phases of quantum spins. Numerical simulations using the density matrix renormalization group (DMRG) method~\cite{YanScience, JiangNatPhys, DepenbrockPRL2012, ZhuPRB2015,WJHuTriZ2SL2015} have provided evidences that the nearest-neighbor AF Heisenberg model on the kagome lattice has a gapped $\mathbb Z_2$ spin liquid ground state. Moreover, recent experiments~\cite{YSLeeNMR_aps, HanSLCorrImpurity} on the kagome lattice material herbertsmithite suggest that there is a spin gap. Although neither the numerical nor the experimental evidences are spotless~\cite{IqbalKagome2011, CommentOfHanSLCorrImpurity}, we still use the kagome lattice as the main example to demonstrate our method, which can be readily generalized to other lattices.

The rest of the paper is organized as follows. In
Sec.~\ref{sec:sfz2}, we briefly review the concept of symmetry fractionalization and explain its mathematical description, which will be used throughout this paper. In Sec.~\ref{sec:phys-constr}, we derive the key result of this work, which is a constraint that relates the symmetry fractionalization class of the spinon to the projective symmetry representation of the physical degrees of freedom in a unit cell. In Sec.~\ref{sec:afdc}, we derive the relation between anyonic flux density created by a symmetry operation, and the fractionalization of commutation relations between this symmetry operation and translations. This relation is needed in applying the constraint obtained in Sec.~\ref{sec:phys-constr}. In Sec.~\ref{sec:appl-kaomge}, this constraint is applied to the triangular/kagome lattices to obtain a full classification of symmetric gapped $\mathbb Z_2$ spin liquid states on such lattices.

\section{Symmetry fractionalization in $\mathbb{Z}_2$ spin liquids}
\label{sec:sfz2}

In this section, we briefly review the physical phenomena of symmetry
fractionalization in the
context of gapped 2D $\mathbb Z_2$ spin liquids~\cite{EssinPRB2014, BarkeshliX}. We
are interested in symmetric $\mathbb{Z}_2$ spin liquids realized in lattice
models of spin-$1/2$ magnets. The largest symmetry group of such
lattice models is $\mathcal{G}=G\times G_s$ with $G_s$ being the space
group. The ``on-site'' part of the symmetry group is
$G=\mathrm{SO}(3)\times \mathbb{Z}_2^T$, where SO(3) denotes the
spin-rotational symmetry group, and $\mathbb Z_2^T$ denotes the
time-reversal symmetry group. For concreteness, we define the space group operations to
only change the positions of the lattice spins: namely, for
$\mb{g}\in G_s$, the symmetry operation $R_{\mb g}$ acts on the spin
at $\bm r$ as
$R_\mb{g}\bm S_{\bm r}R_\mb{g}^{-1}=\bm{S}_{\mb{g}(\bm{r})}$, without
any actions on the internal degrees of freedom.

The concept of symmetry fractionalization refers to the phenomena that
in a 2D topologically ordered state, nontrivial anyon excitations can
carry fractionalized symmetry quantum numbers, and more generally,
projective representations of the symmetry group $G$~\footnote{More generally, the symmetry operation may also permute anyon types. For example, a $\mathbb Z_2$ symmetry may map an $e$ anyon to an $m$ anyon, in a toric code topological order. However, such possibilities are ruled out by our assumption of a time-reversal-invariant spin-$\frac12$ model on a lattice with an odd number of spin-$\frac12$ per unit cell~\cite{ZaletelPRL2015, Cheng_unpub}. In such systems, the $e$ anyon must carry a half-integer spin, and $m$ anyon must carry an integer spin. Therefore, they cannot be interchanged by a symmetry operation due to the incompatible spin-symmetry fractionalization.}. On the contrary,
trivial local excitations must carry ``integer'' quantum numbers, or
linear representations of the symmetry group. Under the assumption of symmetry
localization~\cite{Essin2013}, global symmetry transformations acting
on a state with multiple well-separated anyons $a_1, a_2, \dots$ should factor into
unitaries $U^{a_j}$ localized near the positions of the anyons (for
space group transformations, one also has to move the locations of the
anyons).

Although the global state must form a linear representation of the symmetry group, the local unitaries can obey the group multiplication law
projectively:
\begin{equation}
  \label{eq:proj}
  U^a(\mb{g})U^a(\mb{h})=\omega_a(\mb g, \mb h)U^a(\mb{gh}),
\end{equation}
where each pair of group elements $\mathbf{g,h}$ is associated with a
$\mathrm{U}(1)$ phase $\omega_a(\mathbf{g,h})$. In addition, the
projective phases must satisfy the fusion rules of anyons: if the
anyon type $c$ appears in the fusion of anyons $a$ and $b$, then
$\omega_a\omega_b=\omega_c$. A well-known result from the algebraic theory of anyons~\cite{BarkeshliX} says that all such phase factors consistent with
fusion rules can be expressed using mutual braidings with an Abelian
anyon:
\begin{equation}
  \label{eq:omega-w}
  \omega_a(\mb{g,h})=M_{a, \cohosub{w}(\mb{g,h})},\quad
  \coho{w}(\mb{g,h})\in \mathcal{A}.
\end{equation}
Here, $\mathcal{A}$ is the group of Abelian anyons, and $M_{ab}$ is the mutual braiding statistical phase between anyons $a$ and $b$. Therefore
equivalence classes of the phase factors $\omega_a(\mb g, \mb h)$ consistent
with both the fusion rule and the associativity law of $\mathcal{G}$ are parametrized by
the so-called fractionalization classes
$[\coho w]$, classified by the second group cohomology
$\mathcal H^2[\mathcal{G}, \mathcal A]$~\cite{Chen2014, BarkeshliX, Tarantino_arxiv}.

Notice that we have provided two equivalent descriptions of symmetry fractionalization, in terms of the phase factors $\omega_a(\mb g,\mb h)$ or the anyon-valued fractionalization class $\coho w(\mb g, \mb h)$.
We shall use them interchangeably at different places in
this paper, depending on convenience.

Now we make these abstract definitions more explicit for $\mathbb{Z}_2$ spin liquids. There are four types of anyons:
the trivial excitations $\mathds1$, the bosonic spinon $e$, the bosonic vison $m$ and the fermionic spinon $\epsilon$, with a nontrivial mutual braiding statistics $M_{em}=M_{e\epsilon}=M_{m\epsilon}=-1$.
Under fusion, they form a group
$\mathcal A=\mathbb Z_2\times\mathbb Z_2$, generated by $e$ and $m$. In terms of the fractionalization class $[\coho{w}]$, symmetry fractionalization is
classified by
$\mathcal H^2[\mathcal{G}, \mathcal A] = \mathcal H^2[\mathcal{G}, \mathbb
Z_2\times\mathbb Z_2]=\mathcal{H}^2[\mathcal{G}, \mathbb{Z}_2]^2$.
In terms of the phase factors $\omega_a(\mb g,\mb h)$, the
$\mathbb Z_2$ fusion rules
$e\times e=m\times m=\epsilon\times\epsilon=\mathds1$ constraint them
to be $\pm1$. Therefore, for each type of anyon $a$ the phase factors $\omega_a$ are classified by
$\mathcal H^2[\mathcal{G}, \mathbb Z_2]$. Furthermore, the fusion
rule $e\times m=\epsilon$ implies
$\omega_\epsilon = \omega_e\omega_m$. So the independent phase factors are $\omega_e$ and $\omega_m$, as expected from the group cohomology classification.
\footnote{
We notice that a different convention for projective phase factors is widely used in literature, particularly in the context of PSG classification of Schwinger/Abrikosov fermion mean-field construction~\cite{wenpsg}. With this convention, depending on the symmetry operations involved, an additional ``twist factor'' may need to be included in the relation between $[\omega_e]$, $[\omega_m]$ and $[\omega_\epsilon]$~\cite{Essin2013,LuBFU,QiCSF}. In this work, we follow the definition described in \Ref{CincioCSL2016} to avoid this complication. In the example of triangular/kagome lattices discussed in Sec.~\ref{sec:appl-kaomge}, the symmetry fractionalization quantum numbers that appear in the constraint all have trivial twist factors, and therefore have the same form in these two definitions.
}

We end this section with two familiar examples of symmetry
fractionalization. First, consider the time-reversal symmetry.
In most QSLs, both the bosonic spinon and the fermionic
spinon are Kramers doublets with $T^2=-1$ (or $\omega_e(T, T)=-1$), while the vison has $T^2=+1$ ($\omega_m(T, T)=1$).
According to Eq.~\eqref{eq:omega-w}, the corresponding anyon-valued fractionalization class is $\coho w(T, T)=m$. Our second example, which is crucial to our discussions, is the
fractionalization of the translation symmetries generated by $T_1$ and $T_2$.
As we will see later, QSLs considered in this work all have fractional commutation relation between the two translations
for visons: $T_1T_2=-T_2T_1$ (notice that the translation symmetry group is $\mathbb{Z}\times\mathbb{Z}$). In particular, here we consider spin liquids where both the vison $m$ and the fermionic spinon $\epsilon$ carries $T_1T_2=-T_2T_1$, and the bosonic spinon $e$ carries $T_1T_2=T_2T_1$. To formally capture this kind of fractionalized commutation relation, it is convenient to define the following symbols $\beta_a(\mb g, \mb h)$ and $\coho b(\mb g, \mb h)$:
\begin{equation}
  \label{eq:bgh}
  \beta_a(\mb g, \mb h)=\frac{\omega_a(\mb g,\mb h)}{\omega_a(\mb h, \mb g)},\quad
  \coho b(\mb g, \mb h)=\coho w(\mb g, \mb h)\overline{\coho w(\mb h, \mb g)}.
\end{equation}
This definition only applies to two group elements that commute with each other (we will see an example of a generalized version of the ``commutator'' in Eq.~\eqref{eq:bypx} between non-commuting symmetry operations). Using these notations, the fractionalization of translation symmetry is described by $\beta_e(T_1,T_2)=+1$ and $\beta_m(T_1,T_2)=-1$, or equivalently by $\coho b(T_1,T_2)=e$ [see Eq.~\eqref{eq:omega-w}].

\section{Microscopic constraints on spinon's symmetry fractionalization}
\label{sec:phys-constr}

In this section, we describe constraints on symmetry
fractionalization of spinons in a gapped $\mathbb Z_2$ spin liquid with a net
spinon charge per unit cell. We first briefly review the results of
\Refs{ZaletelPRL2015, Cheng_unpub}, which relate fractional
quantum numbers of spinons to the symmetry representation of the
physical degrees of freedom per unit cell, for on-site symmetry
operations. Then we generalize this relation to point-group symmetry
operations.

We focus on systems with an odd number of spin-$\frac12$ per unit cell,
which include most of the candidate spin liquid systems studied
numerically and experimentally. The Lieb-Schultz-Mattis-Oshikawa-Hastings theorem~\cite{LSM, AffleckLieb, OshikawaLSM, HastingsLSM} guarantees that a gapped ground state in such systems
preserving translation and spin-rotational symmetries must be
topological ordered. Recently, \citet{ZaletelPRL2015} and
\citet{Cheng_unpub} showed that this conclusion can be further
refined: symmetry fractionalization in the topological phase is highly
constrained by symmetry properties of microscopic degrees of
freedom. For example, with $\mathrm{SO}(3)$ symmetry and an odd number
of spin-$1/2$'s per unit cell, it was
established~\cite{ZaletelPRL2015, Cheng_unpub} that there must be a
background anyon charge $b$ in each unit cell that carries
spin-$1/2$. Therefore, $b$ has to be a spinon. The physical meaning of
the background anyon charge $b$ is that when an anyon $a$ is
adiabatically transported around a unit cell, a Berry phase $M_{ab}$
is accumulated. In other words, the translation symmetry is
fractionalized: $T_x^{(a)}T_y^{(a)}=M_{ab}T_y^{(a)}T_x^{(a)}$, or
equivalently, $\coho w(T_x, T_y)=b\neq\mathds1$. For $\mathbb{Z}_2$
QSL, without losing generality we assume that the
$e$ anyon and the $\epsilon$ anyon carry a half-integer
spin
~\footnote{If $e$ carries a half-integer spin, $m$ must carry an integer spin. Otherwise the state must have a non-vanishing quantized $\mathrm{SO}(3)$ spin Hall conductance and breaks time-reversal symmetry explicitly.}.
So we can
have $b=e,\epsilon$. The discussion in this paper generally applies to
both cases.

The existence of a nontrivial background spinon charge puts further
constraints on the symmetry fractionalization of the spinon.
\Ref{Cheng_unpub} shows that for an onsite symmetry group
$G$, such constraints relate the fractionalization class to the symmetry
quantum numbers of the physical degrees of freedom in the unit
cell. Formally, the microscopic degree of freedom in the unit cell
can form a projective representation of the symmetry group, labeled by
$[\nu]\in \mathcal{H}^2[G, \mathrm{U}(1)]$. The following relation is derived in \Ref{Cheng_unpub}:
\begin{equation}
  \label{eq:constraint}
  [\nu]=\rho([\omega_b])\cdot[\tau],
\end{equation}
where $[\tau]$ is a ``twist factor'' determined by certain
symmetry fractionalization class known as the ``anyonic spin-orbit coupling (SOC)'', which describes the fractionalization of commutation relations between translations and other symmetries. The precise form of this additional factor will be given later in this section, and in Sec. \ref{sec:appl-kaomge}, we shall see that for spin liquids on the kagome lattice, this factor is always trivial. Here, $\rho$ is a natural map from $\mathcal H^2[\mathcal G, \mathbb Z_2]$ to $\mathcal H^2[G, \mathrm U(1)]$, which is composed of two steps: first, a cocycle in $\mathcal H^2[\mathcal G, \mathbb Z_2]$ is mapped to a cocycle in $\mathcal H^2[\mathcal G, \mathrm U(1)]$, by replacing the Z2 coefficients with U(1) coefficients +/-1;
second, it is further mapped to a cocycle in $\mathcal H^2[G, \mathrm U(1)]$, by restricting group elements that appear in a cocycle to be elements of $\mathcal G$.

In this work, we generalize their result to also include point group
symmetry $G_\text{pt}$. The triangular and kagome lattices, which we focus on in this paper, share the same space group p6m, which is a semidirect product of the translation symmetry group $G_\text{trans}$, and the point group $G_\text{pt}\simeq G_s/G_\text{trans}$. As shown in Fig.~\ref{fig:symop}, $G_\text{trans}$ is generated by two translations $T_1$ and $T_2$, and $G_\text{pt}$ is generated by two mirror reflections $\mu$ and $\sigma$. In the following we will
 refer to $\tilde{G}=G\times G_\text{pt}$ as the extended onsite symmetry group.
One can always
choose the unit cell to be invariant geometrically under
$G_\text{pt}$. Then the microscopic degrees of freedom in the unit
cell should form a representation of
$\tilde{G}$, classified again by
$[\nu]\in\mathcal{H}^2[\tilde{G}, \mathrm{U}(1)]$. We will
show that Eq. \eqref{eq:constraint} should continue to hold in this
more general setting.

The derivation of Eq.~\eqref{eq:constraint} in \Ref{Cheng_unpub} is based on the observation that the 2D lattice system can be viewed as
surface of a three-dimensional weak SPT state,
and then the constraint on symmetry fractionalization is obtained from bulk-boundary correspondence. Unfortunately,
we do not know how to generalize this approach to include point-group
symmetries, since the bulk-surface correspondence of 3D SPT states protected by
point-group symmetries is not well-understood (however, see \Ref{Song_unpub} for a recent development). Instead, here we give a
purely two-dimensional argument, which not only reproduces
the result of \Ref{Cheng_unpub} for onsite symmetries, but also
generalizes to point-group symmetries.

We consider a (simply-connected) region $A$ that contains $n$ unit cells (thus $n$ background spinons), where
$n$ is an odd number. The ground-state wave function can
be Schmidt decomposed with respect to the cut $\partial A$:
\begin{equation}
  \label{eq:schmidt}
  |\Psi\rangle=\sum_\alpha\lambda_\alpha
  |\psi_A\rangle_\alpha\otimes|\psi_{\bar{A}}\rangle_\alpha.
\end{equation}
Here $\bar{A}$ is the complement of $A$.
In the Schmidt decomposition, the Hilbert space $\mathcal H$ is
decomposed into $\mathcal H=\mathcal H_A\otimes\mathcal H_{\bar{A}}$, where
$\mathcal H_{A/\bar{A}}$ are supported on the interior and the exterior of
the cut $\partial A$, respectively. $|\psi_A\rangle_\alpha\in\mathcal H_A$ and
$|\psi_{\bar{A}}\rangle_\alpha\in\mathcal H_{\bar{A}}$ are Schmidt eigenstates in the
two subspaces. Since $\{|\psi_{A}\rangle_\alpha\}$ form a complete
orthonormal basis of $\mathcal H_A$, one can define the
$\tilde G$-representation in the subspace $\mathcal H_A$ using the
basis of Schmidt eigenstates,
\begin{equation}
  \label{eq:Ug}
  R_\mathbf{g}|\psi_A\rangle_\alpha=
  \sum_{\beta}U_{\alpha\beta}(\mathbf{g})|\psi_A\rangle_\beta,
  \quad \mb{g}\in \tilde G.
\end{equation}
Here $R_\mathbf{g}$ should be understood as the restriction of the
global $\mathbf{g}$ transformation to the interior of $A$. The matrices $U(\mathbf{g})$ form a projective representation of $\tilde{G}$:
\begin{equation}
U(\mathbf{g})U(\mathbf{h})=\omega_U(\mathbf{g,h})U(\mathbf{gh}),
	\label{}
\end{equation}
with a U(1)
phase ambiguity.
Hence, different projective representations are classified by
$\mathcal{H}^2[\tilde G, \mathrm{U}(1)]$.

Our argument now roughly proceeds as follows: we will first analyze the symmetry action in a region in terms of the topological degrees of freedom, and then match with the microscopic description of the same symmetry action.

On one hand, we notice that the group $\tilde{G}$ acts on the Hilbert
space $\mathcal H_A$, to which the Schmidt eigenstates
$|\psi_A\rangle_\alpha$ belong. The action of $\tilde{G}$ on
$\mathcal H_A$ is determined by the symmetry transformation of the
physical degrees of freedom, which is a tensor product of the
microscopic constitute on each site. Correspondingly, the factor set $\omega_U$ is given by
\begin{equation}
  \label{eq:uphys}
  [\omega_U]=([\nu])^n.
\end{equation}

On the other hand, the projective representation $U(\mb g)$ also
encodes symmetry actions on the anyonic excitations. For our purpose, it is of crucial importance to understand the interplay of translation symmetry with the extended symmetry group $\tilde{G}$, which naturally come in two forms:
1) Fractionalization of translation symmetries manifests as background anyon charges (i.e. background spinons in our case). $\tilde{G}$ can act projectively on the background spinons. 2) The commutation relations between the translation symmetries and other symmetry transformations from $\tilde{G}$ can fractionalize. This phenomena is dubbed ``anyonic spin-orbit coupling (SOC)'' in \Ref{Cheng_unpub}.

First, we study the symmetry action on the background
anyon charges. Intuitively,  the
entire region $A$ contains a total $b^n$ anyon charge.  Under the
assumption of symmetry localization~\cite{Essin2013}, $R_\mb{g}$ should act projectively
on $b^n$. Denoting the local symmetry transformations by $U_\mb{g}^{b^n}$, they satisfy
\begin{equation}
  \label{eq:bnproj}
  U^{b^n}(\mb g)U^{b^n}(\mb h)
  =[\omega_{b^n}(\mb g, \mb h)]^nU^{b^n}(\mb{gh}).
\end{equation}

Secondly, the symmetry action can change the Schmidt state in a more subtle way: in the presence of anyonic SOC, symmetry action in $A$ can create an ``anyonic flux density'' running across $A$, which in turn creates anyon charge densities on the boundary of $A$, but does not affect the local density matrix inside $A$ at all. This anyonic flux density creation contributes additional
phase factors to $\omega_U(\mb g, \mb h)$.

\begin{figure}[t!]
  \centering
  %\begin{tabular}{cc}
    \subfigure[\label{fig:afdc:flux}]{\includegraphics{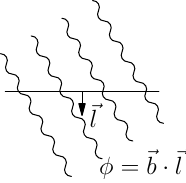}}
    \subfigure[\label{fig:afdc:edge}]{\includegraphics{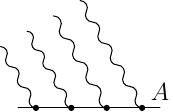}}
    %\subfigure[\label{fig:afdc:str}]{\includegraphics{anyon_soc_lax}}\\
      \subfigure[\label{fig:afdc:xy}]{\includegraphics{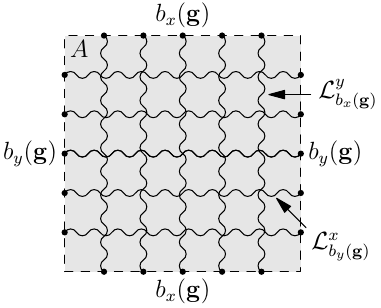}}
    %\end{tabular}
  \caption{Illustrations of the effect of anyon SOC.}
  \label{fig:afdc}
\end{figure}

In general, the anyonic flux density created by a symmetry operation $\mb g$ can be represented using a ``vector'' $\vec b(\mb g)=b_x(\mb g)\hat x+b_y(\mb g)\hat y$. Here, the components of $\vec b(\mb g)$ are Abelian anyon charges $b_{x,y}(\mb g)\in\mathcal A$. Such an anyon flux can be represented by a collection of anyonic string operators, as shown in Fig.~\ref{fig:afdc:flux}. The total anyonic charge carried by these string operators going through $\vec l$ is $\phi=\vec b\cdot\vec l$. In particular, these string operators terminate at the boundary of the region $A$, leaving an anyonic charge density, as shown in Fig.~\ref{fig:afdc:edge}.
In $\mathbb Z_2$ spin liquids, these background string operators can be thought as electric and magnetic field lines.

The value of the vector $\vec b(\mb g)$ is determined by the fractionalization of commutation relations between $\mb g$ and translational symmetries. We leave the computation of $\vec b(\mb g)$ to Sec.~\ref{sec:afdc}, and examine now how the creation of an anyonic flux density $\vec b(\mb g)$ contributes the extra phase factor $\tau$ to Eq.~\eqref{eq:constraint}.
We compute $\tau$ for the example of a rectangular region, with $n_{x,y}$ unit cells in $x$ and $y$ directions, respectively, as shown in Fig.~\ref{fig:afdc:xy}. The symmetry action creates an $n_x$ number of anyon string operators of the type $b_y(\mb g)$ along the $y$ direction, and an $n_y$ number of $b_x(\mb g)$ anyon strings along the $x$ direction, respectively. Hence, the action of $\mathbf g$ not only transforms the anyon within projectively, but also creates anyon string operators. Formally, we write this as
\begin{equation}
  \label{eq:gaction3}
  U(\mathbf g) = U^{b^n}(\mathbf g)
  [\mathcal L_{b_x(\mb g)}^x]^{n_y}
  [\mathcal L_{b_y(\mb g)}^y]^{n_x}.
\end{equation}

When combining the actions of $\mb g$ and $\mb h$, the string
operators contribute an additional phase factor from braiding:
\begin{equation}
  \label{eq:Lphase}
  \begin{split}
  [\mathcal L_{b_x(\mb g)}^x]^{n_y}
  &[\mathcal L_{b_y(\mb g)}^y]^{n_x}
  [\mathcal L_{b_x(\mb h)}^x]^{n_y}
  [\mathcal L_{b_y(\mb h)}^y]^{n_x}\\
  &=M_{b_y(\mb g), \overline{b_x(\mb h)}}^{n_yn_x}
  [\mathcal L_{b_x(\mb{gh})}^x]^{n_y}
  [\mathcal L_{b_y(\mb{gh})}^y]^{n_x}.
  \end{split}
\end{equation}
Combining this phase factor with the cocycle factors in
Eqs.~\eqref{eq:uphys} and \eqref{eq:bnproj}, we obtain
\begin{equation}
  \label{eq:constraint-n}
  \omega_U(\mathbf g, \mathbf h) =
  [\nu(\mathbf g, \mathbf h)]^n=[\omega_b(\mathbf g, \mathbf h)]^n
  [M_{b_y(\mb g), \overline{b_x(\mb h)}}]^n,
\end{equation}
for general $n=n_xn_y$, and this in turn implies that
\begin{equation}
  \label{eq:constraint-soc}
  \nu(\mathbf g, \mathbf h)=\omega_b(\mathbf g, \mathbf h)
  M_{b_y(\mb g), \overline{b_x(\mb h)}}.
\end{equation}
This is the ``anomaly-matching'' constraint given at the beginning of this
section in Eq.~\eqref{eq:constraint}, where the twist factor $[\tau]$ has
the following form,
\begin{equation}
  \label{eq:twist}
  \tau(\mb g, \mb h)=M_{b_y(\mathbf g), \overline{b_x(\mathbf h)}}.
\end{equation}

We notice that when $\mathbf g$ and $\mathbf h$ are onsite symmetries, this relation reproduces the twisted anomaly-matching condition, given by Eq.~(52) of \Ref{Cheng_unpub}, up to a coboundary term. In fact, for onsite symmetries, $\vec b(\mb g)$ is computed in Sec.~\ref{sec:afdc:onsite}, and it is given by Eq.~\eqref{eq:bxy-asoc}.
Hence, Eq.~\eqref{eq:constraint-soc} can be written as
\begin{equation}
  \label{eq:constraint-soc2}
  \begin{split}
  \nu(\mathbf g, \mathbf h)=&\omega_b(\mathbf g, \mathbf h)
  \frac{R^{\cohosub{b}(T_x, \mathbf g), \cohosub{b}(T_y, \mathbf h)}}
  {R^{\cohosub{b}(T_y, \mathbf g),\cohosub{b}(T_x, \mathbf h)}}\\
  &\times\frac{R^{\cohosub{b}(T_y,\mathbf{gh}),\cohosub{b}(T_x,\mathbf{gh})}}
  {R^{\cohosub{b}(T_y, \mathbf g),\cohosub{b}(T_x,\mathbf g)}
  R^{\cohosub{b}(T_y, \mathbf h), \cohosub{b}(T_x,\mathbf h)}},
\end{split}
\end{equation}
where the last term is a $2$-coboundary, and the R symbol $R^{ab}$ denotes the Berry phase associated with exchanging two anyons with charge $a$ and $b$, respectively.
Hence, this anomaly-matching constraint is equivalent to the following:
\begin{equation}
  \label{eq:meng-52}
  \nu(\mathbf g, \mathbf h)=\omega_b(\mathbf g, \mathbf h)
  \frac{R^{\cohosub{b}(T_x, \mathbf g), \cohosub{b}(T_y, \mathbf h)}}
  {R^{\cohosub{b}(T_y, \mathbf g),\cohosub{b}(T_x, \mathbf h)}},
\end{equation}
which reproduces Eq.~(52) of \Ref{Cheng_unpub}.

\section{Anyonic flux density created by symmetry operations}
\label{sec:afdc}

In this section, we study the anyonic flux density created by a symmetry operation $\mb g\in\tilde G$. In particular, we compute the vector $\vec b(\mb g)$ for both onsite and point-group symmetry operations, and show that the result can be related to the fractionalization of commutation relation between $\mb g$ and translational symmetries, known as the anyonic SOC when $\mb g$ is onsite. The results of this section are not only useful for deriving the extra phase factor $[\tau]$ in the constraint of Eq.~\eqref{eq:constraint}, but also provide physical and observable effects of these symmetry fractionalization quantum numbers. Consequently, the creation of $\vec b(\mb g)$ can be used as a way to define and to measure these symmetry fractionalization quantum number.

\begin{figure}[t!]
  \centering
  %\begin{tabular}{cc}
    \subfigure[\label{fig:asoc:g}]{\includegraphics{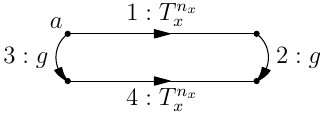}}
    \subfigure[\label{fig:asoc:mirror}]{\includegraphics{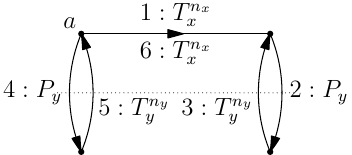}}\\
    %\end{tabular}
  \caption{Illustrations of the effect of anyon SOC.}
  \label{fig:asoc}
\end{figure}

In the following we first review the relation between $\vec b(\mb g)$ and anyonic SOC, for onsite symmetries, which is discussed in details in \Ref{Cheng_unpub}, and then study the generalization to point-group symmetries. The general strategy is that we design processes that compute the braiding phases $M_{a, b_{x,y}(\mb g)}$. The flux $\vec b(\mb g)$ is then determined from the braidinging phases.

\subsection{Onsite symmetries}
\label{sec:afdc:onsite}

First, we consider an onsite symmetry operation $\mb g$, which can be
either unitary or anti-unitary (e.g. time-reversal symmetry). With anyonic SOC, $\mb{g}$ action on the ground state effectively creates an anyonic flux density $\vec b(\mb g)$. Such a flux density $\vec b$ can be determined by moving a test anyon $a$ along a vector $\vec l=l_x \hat{x}+l_y\hat{y}$. The commutation between the $a$ string and the background string operators results in a phase $M_{a,\phi}$ where $\phi=\vec b\cdot\vec l$. Using this method, we will measure the anyonic flux density creation with the following thought experiment: consider creating a pair of anyons $a$ and $\bar a$ from the vacuum, and translate $a$ by $n_x$ unit lengths in the $x$-direction by applying a string operator $\mathcal L_a^x$. This can be thought as applying $T_{x}^{n_x}$ to $a$.  We then apply the $\mb g$ symmetry transformation to the state. The sequence is illustrated in Fig. \ref{fig:asoc:g} as $1\rightarrow 2$. We can also switch the order of translating $a$ and the $\mb{g}$ action, corresponding to $3\rightarrow 4$ in Fig. \ref{fig:asoc:g}. Via the assumption of symmetry localization, it is easy to see that the Berry phase in this process is given by the commutator of $T_{x}^{n_x}$ and $\mb{g}$, and by definition is equal to $[\beta_a(T_{x}, \mb{g})]^{n_x}$. Using Eq. \eqref{eq:omega-w}, this phase can be expressed as
\begin{equation}
  \label{eq:mab-ww}
  [\beta_a(T_{x},\mb g)]^{n_x} = [M_{a,\cohosub{b}(T_{x},\mb{g})}]^{n_x}.
\end{equation}
Since the braiding phase $M_{a,c}$ is also the commutator of two crossing string operators moving $a$ and $c$ respectively, one can re-interpret Eq. \eqref{eq:mab-ww} in the following manner: first, the $\mb{g}$ action creates a string $\mathcal{L}_{b_y(\mb g)}^y$ along $y$ per unit length in the $x$-direction. The phase in Eq. \eqref{eq:mab-ww} then results from the nontrivial commutator between $\mathcal{L}_{a}^x$ and $\mathcal{L}_{b_x(\mb{g})}$ (the $\mathcal{L}_a^x$ string have to cross $n_x$ $\mathcal{L}_y^a$ strings):
\begin{equation}
  \label{eq:llnx}
  \mathcal L_a^x[\mathcal L_{b_x(\mb{g})}^y]^{n_x}
  =[M_{a,b_x(\mb{g})}]^{n_x}[\mathcal L_{b_x(\mb{g})}^y]^{n_x}\mathcal L_a^x.
\end{equation}
This is illustrated in Fig. \ref{fig:asoc:g}. Comparing the Berry phases in Eqs.~\eqref{eq:mab-ww} and \eqref{eq:llnx}, we conclude that $b_y(\mb g)=\coho b(T_x, g)$~\footnote{In a two-dimensional topological order, the braiding is always nondegenerate: $M_{a,b}=M_{a,c}$ for all $a$ if and only if $b=c$.}. The other component $b_x(\mb g)$ can be computed similarly. Finally, we get the relation between $\vec b(\mb g)$ and the anyonic SOC,
\begin{equation}
  \label{eq:bxy-asoc}
  b_y(\mb g)=\coho b(T_x,\mb g),\quad
  b_x(\mb g)=\overline{\coho b(T_y,\mb g)}.
\end{equation}

\subsection{Point-group symmetries}
\label{sec:afdc:gpt}

Next, we show that point-group symmetries have a similar effect of
creating anyonic flux densities associated with the fractionalization of their commutation relations with translations. Here, we only discuss mirror reflections, since other point-group symmetry operations can be reduced to them (as shown in Fig.~\ref{fig:symop}, the point group is generated by two mirror reflections). Without losing generality, we describe our results in the context of mirror reflections on a square lattice. For simplicity, for the rest of this section we restrict ourselves to the case of a $\mathbb Z_2$ (toric code) topological order.

We first consider a mirror symmetry $P_y$, which maps $y$ to $-y$. The anyonic
flux created by $P_y$ in the $y$ direction, which is perpendicular to
the mirror axis, can also be related to the fractionalization of the commutation relation between $T_x$ and $P_y$. To detect this flux density, we draw a test
string operator along the mirror axis. The Berry phase of exchanging
this test string operator $\mathcal L_a^x$ with the anyonic flux
$[\mathcal L_{b_y(P_y)}^y]^{n_x}$, can be interpreted as the Berry
phase of exchanging two sequences of symmetry actions shown in Fig.~\ref{fig:asoc:mirror}: The first sequence is applying $L_a^x$ and then $P_y$, which maps to translating an $a$ anyon by $T_x^{n_x}$ followed by applying $P_y$, and finally applying $T_y^{n_y}$ to move the anyon back to the end of the string operator. In Fig.~\ref{fig:asoc:mirror} this sequence is $1\rightarrow 2\rightarrow 3$. The second sequence is applying $P_y$ and then $L_a^x$, which maps to first applying $P_y$, then applying $T_y^{n_y}$ to move the anyon back to the beginning of the string operator, and finally apply $T_x^{n_x}$. This is illustrated as $4\rightarrow 5\rightarrow 6$ in Fig.~\ref{fig:asoc:mirror}. Here, additional translations in the $y$ direction ($T_y^{n_y}$) are applied to ensure that the string operator $\mathcal L_a^x$ appears at the same location in the two sequences. The Berry phase can be computed as the following,
\begin{equation}
  \label{eq:}
  \begin{split}
  T_x^{n_x}T_y^{n_y}P_ya
  &=[\omega_a(T_x,T_y)]^{n_xn_y}T_y^{n_y}T_x^{n_x}P_ya\\
  &=[\omega_a(T_x,T_y)]^{n_xn_y}[\omega_a(T_x,P_y)]^{n_x}
  T_y^{n_y}P_yT_x^{n_x}a.
  \end{split}
\end{equation}
The two interpretations of the same Berry phase should be equated:
\begin{equation}
	{M}_{a, b_y(P_y)}=[\omega_a(T_x,T_y)]^{n_y}\omega_a(T_x, P_y).
	\label{eqn:byphase}
\end{equation}
 Formally we should have
\begin{equation}
  \label{eq:bypyny}
  b_y(P_y)=\coho b(T_x, P_y)[\coho b(T_x,T_y)]^{n_y}.
\end{equation}
When $\coho b(T_x,T_y)\neq\mathds1$, this equation only gives consistent results if $n_y$ has a fixed parity.

Let us go back to Eq. \eqref{eqn:byphase}. To determine $b_y(P_y)$, we need at least two test anyons, which we will choose to be the background spinon $b=\coho{b}(T_x, T_y)$ and the vison $v$. For the background spinon, $\omega_b(T_x,T_y)=1$ so the choice of $n_y$ does not matter. If the test anyon is a vison, we now argue that the parity of $n_y$ only depends on the geometric property of the mirror axis. For example, consider the site-centered reflection $P_y$. To see why, consider moving a vison by $T_y^{-n_y}T_x^{-n_x}T_y^{n_y}T_x^{n_x}$, so that the path is symmetric under $P_y$. $n_xn_y$ unit cells are enclosed by the path and the Berry phase is $(-1)^{n_xn_y}$. On the other hand, since the area enclosed by the path is symmetric under $P_y$, it is obvious that the number of physical spins inside the area must be $n_x$ mod $2$. We thus conclude that $(-1)^{n_xn_y}=(-1)^{n_x}$, which implies $n_y$ is odd. Combining these arguments, we can unambiguously conclude that
\begin{equation}
	b_y(P_y)=\coho{b}(T_x, P_y)\coho{b}(T_x,T_y).
	\label{eq:bypy}
\end{equation}
 We also notice that site-centered $P_y$ is needed for our general proof to work, since there must be an odd number of spin-$1/2$'s in the region $A$.

Similarly, $b_y(P_x)$ is related to the symmetry fractionalization associated with $T_x$ and $P_x$. Because $P_x$ and $T_x$ do not commute, but instead satisfy a twisted commutation relation $T_xP_x=P_xT_x^{-1}$, we can not naively apply the previous definition of commutation relation fractionalization. As we show below, $b_y(P_x)$ is determined by a twisted version of commutation relation fractionalization,
\begin{equation}
  \label{eq:bypx}
  b_y(P_x)=\coho w(T_x, P_x)\coho w(T_x,T_x^{-1})
    \overline{\coho w(P_x, T_x^{-1})}.
\end{equation}
Similar conclusions hold for $P_y, T_y$ as well as the inversion $I$ with $T_{x,y}$.

\begin{figure}[t!]
  \centering
  \includegraphics[width=\columnwidth]{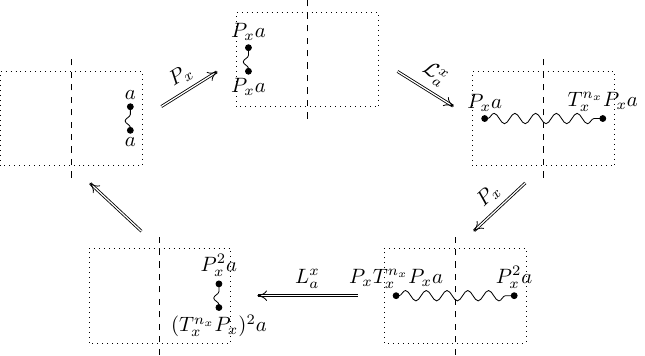}
  \caption{Illustration of the physical process determining the anyonic flux creation $b_y(P_x)$.}
  \label{fig:txpx}
\end{figure}

The anyonic flux density $b_y(P_x)$ can be computed from the Berry phase associated with applying a test string operator $L_a^x$ and $P_x$ in opposite orders. In particular, the Berry phase $M_{a,b_y(P_x)}$ is obtained through the following operations: $(\mathcal L_a^x)^{-1}P_x^{-1}\mathcal L_a^xP_x=\mathcal L_a^xP_x\mathcal L_a^xP_x$. This Berry phase can be related to symmetry fractionalization, by viewing these operations as symmetry actions on test anyon charges. As shown in Fig.~\ref{fig:txpx}, we first create two test charges of type $a$ from the vacuum, to the right of the mirror axis, and denote the initial state by $a\times a$. The operation $P_x$ reflects both anyons to the left, and we denote the result by $P_xa\times P_xa$. Next, the test string $\mathcal L_a^x$ is created by translating one of the anyon to the right, and the state becomes $P_xa\times T_xP_xa$. Next, $P_x$ maps them to $P_xT_xP_xa\times P_x^2a$. Finally, applying the test string operator moves the anyon on the left back to the right, and we get $T_xP_xT_xP_xa\times P_x^2a$. Comparing this to the initial state, we conclude that the Berry phase accumulated is $\omega^a(T_xP_x, T_xP_x)\omega^a(P_x, P_x) = M_{a,b_y(P_x)}$. Hence, we conclude that
\begin{equation}
  \label{eq:bypx2}
  b_y(P_x) = \coho w(T_xP_x, T_xP_x)\coho w(P_x, P_x).
\end{equation}
Using the cocycle equation $\coho w(\mb g, \mb h)\coho w(\mb g\mb h, \mb k) =\coho w(\mb h, \mb k)\coho w(\mb g, \mb{hk})$, one can rewrite this result in the form of Eq.~\eqref{eq:bypx}.

We notice that to determine the anyonic flux density created by a
mirror symmetry operation, it is more convenient to decompose the
flux density $\vec b$ into components in directions parallel and
perpendicular to the mirror axis, as in our discussion above, because
in such a setup, the test string is invariant under the mirror
symmetry. In the contrary, the flux density creation associated with
onsite and inversion symmetries can be determined using decompositions
in any directions.

Besides determining the twist factor $[\tau]$ in Eq.~\eqref{eq:twist}, the effect of anyonic flux density creation can also be used in defining and detecting the (twisted) commutation relation fractionalization between translations and elements of $\tilde G$. The Berry phase computations we have presented can serve as physical definitions of these fractionalization classes, and the anyonic flux creation provides an intuitive picture for the physics.

The effect of anyonic flux density creation can also be exploited in numerical diagonsis of the corresponding symmetry fractionalization classes. The fractionalization of commutation relations between one translation and another (unitary) symmetry operation $\mb g$ can be measured as the difference in eigenvalues of $\mb g$ operators on ground states in different topological sectors~\cite{QiCSF,ZLVPSG}, with appropriate lattice geometry. This result can be readily reproduced using anyonic flux density creation. For simplicity, we consider the example of an onsite unitary symmetry $\mb g$.  The symmetry operation $\mb g$ creates anyonic flux along the $y$ direction, which can be represented as string operators $[\mathcal L_{b_y(\mb g)}^y]^{n_x}$. On a minimally entangled state (MES) on a torus, with an anyonic flux $a$ going through along the $x$ direction, these string operators contributes a nontrivial phase factor $[M_{a,b_y(\mb g)}]^{n_x}=[\beta_a(T_x, \mb g)]^{n_x}$. Thus, if $n_x$ is odd, the $\mb g$ eigenvalues of different MESs encode the commutation relation fractionalization $\beta_a(T_x,\mb g)$.

\section{Application to gapped spin liquids on the kagome/triangular lattice}
\label{sec:appl-kaomge}

In this section, we apply the constraint Eq.~\eqref{eq:constraint}
to gapped spin liquids in spin-$\frac12$ models on the kagome/triangular
lattice. This constraint greatly reduces the number of possible
symmetry fractionalization classes on spinons. Together with previously
obtained constraints on symmetry fractionalization of
visons~\cite{QiSFV2015X}, we obtain a complete classification of
gapped symmetric $\mathbb Z_2$ spin liquids.

\begin{figure}
  \centering
  \subfigure[\label{fig:symop:kagome}Kagome lattice]
  {\includegraphics{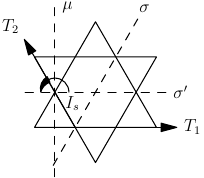}}
  \subfigure[\label{fig:symop:tri}Triangular lattice]
  {\includegraphics{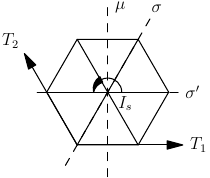}}
  \caption{Generators of the space group of the kagome/triangular lattice, $G_s=p6m$.}
  \label{fig:symop}
\end{figure}

To begin, we recall that the space group of the triangular/kagome lattice,
$G_s=p6m$, is an extension of the translational symmetry group $G_\text{trans}$,
generated by $T_1$ and $T_2$, by the point group $G_{\text{pt}}=C_{6v}$,
generated by two mirror reflections $\mu$ and $\sigma$, as shown in
Fig.~\ref{fig:symop}. It is also convenient to define a site-centered inversion $I_s$, as shown in Fig.~\ref{fig:symop}, and the mirror reflection with the mirror axis along the $T_1$ direction $\sigma^\prime=I_s\mu$. Here, we choose two mirror symmetries $\mu$ and $\sigma^\prime$, whose axes intersect at a lattice site (consequently, the product of mirror reflections gives site-centered rotations), such that a region symmetric under two mirror reflections contain an even number of unit cells.
We notice that the relation between $\sigma^\prime$ (and also $I_s$) and the generators of $p6m$ is different for the two different lattices, although in both cases we have defined $I_s=\mu\sigma'$. In terms of notations introduced in Sec.~\ref{sec:sfz2} and Sec.~\ref{sec:phys-constr}, the total symmetry group is $\mathcal G=\textrm{SO}_3\times\mathbb Z_2^T\times G_s=\textrm{SO}_3\times\mathbb Z_2^T\times p6m$,
and the extended onsite symmetry group is $\tilde G=\textrm{SO}_3\times\mathbb Z_2^T\times G_{\text{pt}}=\textrm{SO}_3\times\mathbb Z_2^T\times C_{6v}$.

To classify symmetry fractionalizations we first compute the second group cohomology
$\mathcal{H}^2[\mathcal{G}, \mathbb Z_2]=\mathbb Z_2^8$. The $2^8=256$ different cohomology classes are labeled by eight $\mathbb Z_2$ invariants:
\begin{equation}
  \label{eq:omega}
  (\omega^{\text{spin}}, \omega^{12},
  \omega^\mu, \omega^\sigma, \omega^{I_s}, \omega^T, \omega^{\mu
    T}, \omega^{\sigma T}),
\end{equation}
where $\omega^{\text{spin}}=0$ or $\frac12$ denotes whether the anyon
carries an integer or a half-integer spin, respectively;
$\omega^{12}=\beta(T_1,T_2)=\pm1$; the other six variables are all given
in the form of $\omega^X=\omega(X, X)=\pm1$, denoting the fractional
quantum number $X^2=\pm1$.   As we have already
discussed in Sec.~\ref{sec:sfz2}, we need to specify $[\omega_e]$
and $[\omega_m]$.

The fact that $\omega^e_\text{spin}=1/2$ and $\coho{b}(T_x, T_y)=e$
 completely determines the symmetry
fractionalization of the vison in a gapped $\mathbb Z_2$ spin
liquid~\cite{QiSFV2015X}. First, the background spinon charge density
implies that the $T_1$ and $T_2$ anti-commute when acting on a vison, i.e. $\beta_m(T_1,T_2)=-1$.
Intuitively, a vison acquires a Berry phase of $\pi$
when it is adiabatically transported around a unit cell, because of the braiding with the background spinon charge.
All the other fractional quantum numbers can be constrained using the method of flux-fusion
anomaly test~\cite{HermeleFFAT, QiSFV2015X}. To summarize, the symmetry
fractionalization of visons is fixed:
\begin{equation}
  \label{eq:omega-m}
  [\omega_m]=(0, -1, +1, +1, -1, +1, +1, +1).
\end{equation}
We notice that, for the convenience of applying the constraint in Eq.~\eqref{eq:constraint}, we consider a site-centered inversion $I_s$, instead of a plaquette-centered inversion $I_p$ as in \Ref{QiSFV2015X}. As a result, the vison carries $I_s^2=-1$. As discussed in \Ref{QiSFV2015X}, this is equivalent to $I_p^2=+1$~\footnote{The fractional quantum number $\omega^{I_p}$, labeling $I_p^2=\pm1$, is not an independent quantum number. Instead, it can be expressed using the fractional quantum numbers defined in Eq.~\eqref{eq:omega} as $\omega^{I_p}=\omega^{12}\omega^{I_s}$. Furthermore, for spin liquids where vison has $\omega^{12}=-1$, this implies that $\omega^{I_p}=-\omega^{I_s}$.}, and the vison PSG specified in Eq.~\eqref{eq:omega-m} is the same as the odd Ising gauge theory on the kagome/triangular lattice~\cite{SachdevFFIM1999, SenthilZ2SL2000, MoessnerTri2001, MoessnerZ2Gauge2001}. We also notice that, although the fractional quantum numbers in Eq.~\eqref{eq:omega-m} appear to be the same for vison PSGs on the kagome and triangular lattices, the two vison PSGs are indeed different~\footnote{We thank Michael Hermele for discussions on this point.}. This is because these quantum numbers are defined in terms of symmetry operations $I_s$, which has different forms on the two types of lattices, in terms of the generators of the $p6m$ space group.

Now we turn to the symmetry fractionalization of spinons. To apply Eq. \eqref{eq:constraint} we first study the symmetry representations of physical degrees of freedom in a unit
cell. As discussed in Sec.~\ref{sec:phys-constr},  they are classified by the  $\mathcal{H}^2[\tilde G, \mathrm U(1)]=\mathbb Z_2^5$, where the extended onsite symmetry group $\tilde G=\mathrm{SO}_3\times\mathbb Z_2^T\times C_{6v}$. These comology classes are labeled by the following invariants:
\begin{equation}
  \label{eq:ophys}
  (\omega^{\text{spin}}, \omega^T, \omega^{\mu T},
  \omega^{\sigma T}, \omega^{\mu,\sigma'}).
\end{equation}
The first four invariants already appear in Eq.~\eqref{eq:omega} [i.e. they are actually $\mathrm{U}(1)$ invariants], and the last one
$\omega^{\mu,\sigma^\prime}=\beta(\mu,\sigma^\prime)=\pm1$ represents the commutation relation
fractionalization between $\mu$ and $\sigma^\prime$. For spin-$1/2$ magnets on the triangular/kagome lattices, it is straightforward to check that the invariants are given by $[\omega_{\text{phys}}]=(\frac12,-1, -1, -1, +1)$.

%It is very useful to notice that there is a natural embedding of $\mathcal{H}^2[\tilde G, \mathrm U(1)]$ into
%$\mathcal{H}^2[\tilde{G}, \mathbb Z_2]$. Namely, a nontrivial cohomology class of $\tilde{G}$ with $\mathrm{U}(1)$ coefficients is necessarily a nontrivial class with $\mathbb{Z}_2$ coefficients.
The natural map $\rho$ between $\mathcal{H}^2[\tilde{G}, \mathbb Z_2]$ and $\mathcal{H}^2[\tilde G, \mathrm U(1)]$ can be understood intuitively.
We already mentioned that $\omega^{\text{spin}}, \omega^T, \omega^{\mu T}, \omega^{\sigma T}$ are identified between the two, and $\omega^{\mu,\sigma^\prime}=\omega^{I_s}\omega^\mu\omega^\sigma$.
[This is because the symmetry fractionalization of $I_s^2=(\mu\sigma')^2$ can be factorized as the product of $\mu^2$, $\sigma^{\prime2}$, and the commutation relation between $\mu$ and $\sigma^\prime$.] $\omega^\mu$ and $\omega^{I_s}$ are the only true ``$\mathbb{Z}_2$'' invariants  in $\mathcal{H}^2[\tilde{G}, \mathbb{Z}_2]$.
In summary, using the notations in Eqs.~\eqref{eq:omega} and \eqref{eq:ophys}, the map $\rho$ has the form
\begin{equation}
  \label{eq:rho}
  \rho([\omega]) = (\omega^{\text{spin}}, \omega^T, \omega^{\mu T}, \omega^{\sigma T}, \omega^{I_s}\omega^\mu\omega^\sigma).
\end{equation}

We now apply the constraint in Eq.~\eqref{eq:constraint} to the fractionalization of
the background spinon $b$.
First, we can show that the twist factor $\tau$ in Eq.~\eqref{eq:constraint}
is always trivial. We observe that, among the invariants listed in Eq.~\eqref{eq:omega} that label
different fractionalization classes in $\mathcal H^2[\mathcal{G}, \mathbb Z_2]$, none of them involves translations and $\tilde{G}$ at the same time, i.e. no anyonic SOC appears in the list. In fact, as shown by an explicit computation in Appendix~\ref{sec:asoc-kagome}, all anyonic SOC fractionalization classes $\coho{b}(T_{1,2},\mb{g})$ [including the twisted version in Eq.~\eqref{eq:bypx}], and hence the anyonic flux densities $b_{x,y}(\mb{g})$, are either $\mathds1$ or the background spinon $b=\coho b(T_1, T_2)$ for all $\mb{g}\in \tilde{G}$.
Since in a $\mathbb{Z}_2$ spin liquid, we always have $M_{\mathds1, b}=M_{b,b}=+1$, the twist factor
$[\tau]$ evaluates to $1$ identically.

With $[\tau]$ out of the way, the
constraint in Eq.~\eqref{eq:constraint} states that the symmetry
fractionalization of the background spinon must match the projective
representation of the physical degrees of freedom in each unit cell.
In particular, according to Eq.~\eqref{eq:rho},
it means that among the eight invariants, $\omega_b^{\text{spin}}, \omega_b^T, \omega_b^{\mu T}, \omega_b^{\sigma T}$ are directly fixed by the symmetry properties of the unit cell, and $\omega_b^\sigma$ is not independent. In summary, we establish the following constraints:
\begin{equation}
	\begin{gathered}
	\omega_b^{\text{spin}}=\omega_\text{phys}^{\text{spin}}, \omega_b^\sigma=\omega_\text{phys}^{\mu,\sigma'}\omega^{I_s}_b\omega^\mu_b,\\
\omega_b^X = \omega_\text{phys}^X,  \:X=T, \mu T, \sigma T.
\end{gathered}
	\label{eq:ck}
\end{equation}

The remaining unconstrained ones are just $\omega_b^{12}$, $\omega_b^\mu$ and $\omega_b^{I_s}$, so we have $2^3=8$ classes of spinon PSGs. Since all the vison PSGs are completely fixed, we conclude that there are at most eight distinct types of symmetric $\mathbb{Z}_2$ spin liquid with a background spinon charge. There are actually no further constraints since all eight possible states have
been constructed previously using bosonic~\cite{Wang2007} or fermionic
parton constructions~\cite{YMLuKagomePSG2011, LuBFU, Zheng_arxiv,YMLuTri2015X}, and tensor
product states~\cite{SJiangTPS2015X}. Therefore, we have obtained a complete
classification of gapped symmetric $\mathbb Z_2$ spin liquids in translation-invariant
spin-$\frac12$ kagome/triangular lattice models. The results are summarized in
Table~\ref{tab:fqn}.

\begin{table}[htbp]
  \centering
  \begin{tabular*}{\columnwidth}{@{\extracolsep{\fill}}cccc}
    \hline\hline
    $[\omega]$ & Relations & $[\omega_e]$ & $[\omega_m]$\\
    \hline
    $\omega^{\text{spin}}$ & $(e^{i\pi S^{x,y,z}})^2=\pm1$ & $\frac12$ & 0\\
    $\omega^{12}$ & $T_1T_2=\pm T_2T_1$ & $\pm1$ & $-1$\\
    $\omega^\sigma$ & $\sigma^2=\pm1$ & $\pm1$ & $+1$\\
    $\omega^\mu$ & $\mu^2=\pm1$ & $\omega^\sigma_e\omega^I_e$ & $+1$\\
    $\omega^{I_s}$ & $I_s^2=(\mu\sigma')^2=\pm1$ & $\pm1$ & $-1$\\
    $\omega^T$ & $T^2=\pm1$ & $-1$ & $+1$\\
    $\omega^{\sigma T}$ & $(\sigma T)^2=\pm1$ & $-1$ & $+1$\\
    $\omega^{\mu T}$ & $(\mu T)^2=\pm1$ & $-1$ & $+1$\\
    \hline
  \end{tabular*}
  \caption{Fractional quantum numbers in symmetric gapped $\mathbb
    Z_2$ spin liquids on the kagome lattice.}
  \label{tab:fqn}
\end{table}

\section{Conclusion}
\label{sec:discuss}

In this work, we study the classification of symmetry fractionalization carried by spinons in gapped 2D $\mathbb Z_2$ spin liquids. For gapped spin liquids realized in spin systems with an odd number of spin-$\frac12$ degrees of freedom per unit cell, we generalize the constraint of \citet{Cheng_unpub} to point-group symmetries. The generalized constraint establishes a relation between the symmetry fractionalization of the spinon and the symmetry representation of the physical degrees of freedom in one unit cell. Applying this constraint to triangular/kagome lattices, we obtain a full classification (up to stacking SPT layers) of gapped $\mathbb Z_2$ spin liquids. In particular, we show that the eight spin liquid states constructed previously exhaust all possibilities~\cite{Wang2007,YMLuKagomePSG2011,LuBFU,Zheng_arxiv,YMLuTri2015X,SJiangTPS2015X}.

We notice that the Schwinger/Abrikosov fermion construction actually yields 20 different PSGs for $\mathbb{Z}_2$ spin liquid states on the kagome lattice~\cite{YMLuKagomePSG2011,LuBFU}. Eight of them correspond to symmetric gapped spin liquids in Table \ref{tab:fqn}~\cite{LuBFU}, while the rest, if naively translating the PSG to fractionalization class of the fermionic spinons, violate the constraints in Eq. \eqref{eq:ck}. However, the fermionic mean-field states are always gapless~\footnote{In these mean-field states, the band structures of the fermionic spinons have gapless modes, i. e. nodes or nodal points, which cannot be gapped out without breaking the symmetries.} with these 12 classes of PSG, i.e. they are \emph{gapless} $\mathbb{Z}_2$ spin liquids. Moreover, on other lattices, spin liquids where all anyons carry integer spins and/or $T^2=+1$ [and thus violating the constraint in Eq.~\eqref{eq:constraint}] can also be realized as gapless spin liquids~\cite{wenpsg,BiswasMSL,GChenMSL}. It was known that the gaplessness of fermionic spinons are protected by the PSG at the level of mean-field states, and our
derivation of the constraints in Eq.~\eqref{eq:ck} provides a non-perturbative
understanding of the symmetry-protected gaplessness.

As we mentioned earlier, \Ref{Cheng_unpub} argued that 2D systems with a spin-$1/2$ per unit cell can be
viewed as the surface of a 3D weak SPT state protected by $\mathrm{SO}(3)$ and translation symmetries,
if only physical degrees of freedom carrying linear representations of $\mathrm{SO}(3)$ are
allowed.
The surface of such a 3D SPT state must be anomalous in a precise manner that can be canceled by ``anomaly in-flow'' to the bulk, and a symmetric surface state can either be gapless or gapped with an
intrinsic topological order satisfying the constraint in
Eq.~\eqref{eq:constraint}. We believe such a bulk-surface correspondence can be generalized to point-group symmetries as well, and for gapped symmetry-preserving surface states Eq. \eqref{eq:constraint} can be viewed as the precise form of such correspondence~\cite{Cheng_unpub}. On the other hand, gapless $\mathbb{Z}_2$ spin liquids violating Eq. \eqref{eq:ck} found in Schwinger/Abrikosov fermion constructions should also satisfy such a bulk-surface relation, i.e. they exhibit the correct anomaly. It will be interesting to investigate the relation further, which we will leave for future studies.

We notice that our constraints on the symmetry fractionalization of spinons can be easily adapted to spin models with only $\mathrm{U}(1)_{S_z}$ spin-rotational symmetry [e.g., the $\mathrm{SO}(3)$ symmetry is broken by easy-plane anisotropies], as long as there is still a background ``spinon'' charge per unit cell. Here, a spinon should be understood as an anyon carrying $1/2$ charge under the $\mathrm{U}(1)_{S_z}$ symmetry. However, the constraints on the symmetry fractionalization of visons are much less stringent: the flux-fusion anomaly test only fixes $\omega^T_m, \omega^\mu_m$ and $\omega^\sigma_m$~\cite{HermeleFFAT}, leaving other quantum numbers of visons undetermined. We leave a systematic classification of non-anomalous gapped spin liquids with $\mathrm{U}(1)_{S_z}$ symmetry for future works.
We also notice that our constraints can be straightforwardly generalized to other 2D lattices, like the square lattice. We will also leave this for future works.

\begin{acknowledgements}
  M.C. would like to thank M. Zaletel for many enlightening conservations on related topics. Y.Q. and M.C. are grateful for Michael Hermele for very constructive comments on the manuscript. Y.Q. is supported by the Ministry of Science and Technology of China under Grant No. 2015CB921700. This research was supported in part by Perimeter Institute for Theoretical Physics. Research at Perimeter Institute is supported by the Government of Canada through Industry Canada and by the Province of Ontario through the Ministry of Research and Innovation.

\emph{Note added.} Recently, we were informed of a related work~\cite{LuUnpub}.
\end{acknowledgements}

\appendix

\section{Anyonic spin-orbit coupling on the kagome lattice}
\label{sec:asoc-kagome}

In this appendix, we derive the symmetry fractionalization classes $\coho b(T_{1,2}, \mb g)$, for symmetry operations $\mb g\in\tilde G$ on the kagome lattice. Because of the six-fold rotational symmetry, the second group cohomology $\mathcal H^2[G, \mathbb Z_2]$ factorizes as the following:
\begin{equation}
  \label{eq:h2g-separate}
  \mathcal H^2[\mathcal{G}, \mathbb Z_2]
  =\mathcal H^2[G_{\text{trans}},\mathbb Z_2]\times
  \mathcal H^2[\tilde G, \mathbb Z_2],
\end{equation}
where $\mathcal H^2[G_{\text{trans}},\mathbb Z_2]=\mathbb Z_2$ is labeled by the quantum number $\omega^{12}=\beta(T_1, T_2)$. Therefore, there are no independent quantum numbers describing the anyon SOC, which must then be completely fixed by the value of $\omega^{12}$.

We first show that an onsite symmetry operation $\mathbf g$ must have a trivial anyonic SOC, i.e. $\beta(T_{1,2}, \mb g)=1$. Using the cocycle condition of $\mathcal H^2[\mathcal{G}, \mathbb Z_2]$, one can show that
\begin{equation}
  \label{eq:bghk}
  \beta(\mb{gh}, \mb k) = \beta(\mb g, \mb k)\beta(\mb h,\mb k),
\end{equation}
if both $\mb g$ and $\mb h$ commute with $\mb k$. Since $T_1$ and $T_2$ are related by the $C_6$ rotational symmetry, the commutation-relation fractionalization between $\mathbf g$ and $T_{1,2}$ must be the same: $\beta(T_1,\mb g)=\beta(T_2, g)$. Formally, this can be proved using Eq.~\eqref{eq:bghk} and the relation $C_6^2T_1C_6^{-2}=T_2$. Then, using Eq.~\eqref{eq:bghk} one more time, we get $\beta(T_1T_2,\mb g)=\beta(T_1,\mb g)\beta(T_2,\mb g)=+1$, i. e. $\mb g$ must commute with $T_1T_2$. However, the translational symmetry operation of $T_1T_2$ is also related to both $T_1$ and $T_2$ by the $C_6$ symmetry. Therefore, we obtain $\beta(T_1,\mb{g})=\beta(T_2,\mb{g})=1$. In particular, the anyonic SOC associated with the time-reversal symmetry, $\beta(T_{1,2}, T)$, must always be trivial.

Next, we study the symmetry fractionalization classes $\beta(T_{1,2},\mb g)$ when $\mb{g}$ is a point-group symmetry operation. Instead of deriving them formally using the algebraic relations, here we compute them using an alternative approach, by explicitly constructing matrix representations of all $2^4=16$ cohomology classes in $\mathcal H^2[G_s, \mathbb Z_2]$. A matrix representation is a map $\phi: G\rightarrow \mathrm{GL}(V)$ where $V$ is a finite-dimensional complex vector space. Each projective representation gives a cocycle $\omega$, as the group multiplication is only realized projectively,
\begin{equation}
  \label{eq:phi-w}
  \phi(\mb g)\phi(\mb h)=\omega(\mb g,\mb h)\phi(\mb{gh}),\quad
  \omega(\mb g, \mb h)=\pm1,
\end{equation}
and $\omega$ satisfies the cocycle condition automatically, because matrix multiplications are associative. Therefore, each projective representation belongs to a certain class of $\omega\in\mathcal H^2[G, \mathbb Z_2]$. If we tensor product two projective representations $\phi=\phi_1\otimes\phi_2$, the factor set of $\phi$ is the product of those of $\phi_1$ and $\phi_2$. Therefore, we only need to construct projective representations realizing each of the four root classes in $\mathcal H^2[G_s,\mathbb Z_2]=\mathbb Z_2^4$, which has one of the four invariants $\omega^{12}$, $\omega^\mu$, $\omega^\sigma$ and $\omega^I$ being $-1$ and all the other three being $+1$.

\begin{table}[htbp]
  \begin{tabular*}{\columnwidth}{@{\extracolsep{\fill}}ccccc}
    \hline\hline
    $\omega^\mu$ & $\omega^\sigma$ & $\omega^I$
    & $\phi(\mu)$ & $\phi(\sigma)$ \\
    \hline
    $-1$ & $+1$ & $+1$ & $i\tau^2$ & $\tau^3$ \\
    $+1$ & $-1$ & $+1$ & $\tau^3$ & $i\tau^2$ \\
    $+1$ & $+1$ & $-1$ & $\tau^3$ & $\tau^1$\\
    \hline\hline
  \end{tabular*}
  \caption{Projective representations realizing root comology classes with $\omega^{12}=1$. Here, $\tau^i$ denotes the Pauli matrices.}
  \label{tab:root3}
\end{table}

As discussed in Sec.~\ref{sec:appl-kaomge}, the cohomology classes with $\omega^{12}=1$ can be realized by a representation with $\phi(T_1)=\phi(T_2)=\mathds{1}$. In particular, the three root classes can be realized using projective representations summarized in Table~\ref{tab:root3}. The last root class with $\omega^{12}=-1$ can be realized using a 16-dimensional projective representation. Here, we use $e(i, j)$, $i, j=0,\ldots,3$ to denote the 16 basis vectors, and the projective representation is specified as follows
\begin{equation}
  \label{eq:w12rep}
  \begin{split}
    \phi(T_1)e(i, j) &= e(i+1, j)\\
    \phi(T_2)e(i, j) &= (-1)^i e(i, j+1)\\
    \phi(\mu)e(i, j) &= (-1)^{j(j+1)/2}e(-i+j, j)\\
    \phi(\sigma)e(i, j) &= (-1)^{ij}e(j, i).
  \end{split}
\end{equation}
In these equations, $i$ and $j$ are defined mod $4$, i. e. $e(i, j)=e(i+4, j)=e(i, j+4)$. This projective representation is obtained by rewriting the $(p_1,p_2,p_3)=(1,0,0)$ state in \Ref{Wang2007} as a matrix representation.

Using this projective representation, one can explicitly compute the fractionalization of commutation relations associated with the mirror symmetries $\mu$ and $\sigma$, for the cohomology class $\omega^{12}=-1$. First, we calculate the commutation relations between mirror reflections and translations along the mirror axes:
\begin{equation}
  \label{eq:ms-fc1}
    \beta(T_1T_2, \sigma)=\beta(T_1, \sigma^\prime)
    =\beta(T_1T_2^2,\mu)=-1.
\end{equation}

Since in the other root cohomology classes with $\omega^{12}=+1$ all these phase factors are equal to $+1$, we conclude that,  $\beta(T_1T_2,\sigma)=\beta(T_1,\sigma^\prime)=\beta(T_1T_2^2,\mu)=\beta(T_1,T_2)$ holds for all cohomology classes. This implies the following results in $\mathcal H^2[\mathcal{G}, \mathcal{A}]$:
\begin{equation}
  \label{eq:ms-bfc1}
  \coho b(T_1T_2,\sigma) = \coho b(T_1,\sigma^\prime)
  = \coho b(T_1T_2^2,\mu) = \coho b(T_1, T_2).
\end{equation}
Using Eq.~\eqref{eq:bypy}, we see that no nontrivial anyonic flux densities are created by these mirror symmetries along the mirror axis.

Next, we consider the twisted commutation relation between mirror reflections and translations perpendicular to the mirror axes. Motivated by Eq.~\eqref{eq:bypx}, we define the following quantum number,
\begin{equation}
  \label{eq:beta-twisted}
  \tilde \beta(\mb g,\mb h) = \omega(\mb g, \mb h)\omega(\mb g, \mb h^{-1}\mb g\mb h)/\omega(\mb h, \mb h^{-1}\mb g\mb h).
\end{equation}
It is easy to check that $\tilde\beta(\mb g, \mb h)$ is a coboundary-independent invariant of a cohomology class $[\omega]\in\mathcal H^2[\mathcal{G}, \mathbb Z_2]$ for any two group elements. Using the explicit form of the projective representation in Eq.~\eqref{eq:w12rep}, we can show that all the twisted commutation relations have trivial fractionalization,
\begin{equation}
  \label{eq:ms-fc2}
  \tilde\beta(T_1^{-1}T_2, \sigma)=\tilde\beta(T_1T_2^2, \sigma^\prime)=\tilde\beta(T_1,\mu)=+1.
\end{equation}
Hence, for all cohomology classes in $\mathcal H^2[\mathcal{G},\mathcal A]$, the results of Eq.~\eqref{eq:bypx} are trivial. In other words, the mirror symmetries do not create any anyonic flux density in the direction perpendicular to the mirror axes. Therefore, the mirror reflections $\mu$, $\sigma$ and $\sigma^\prime$ do not create any nontrivial anyonic flux density. As a result, we can safely ignore the $[\tau]$ factor, when applying the constraint in Eq.~\eqref{eq:constraint} to the triangular/kagome lattices in Sec.~\ref{sec:appl-kaomge}.

\bibliography{psg,lu}
\end{document}